\documentclass[aps,prl,twocolumn,superscriptaddress,showpacs]{revtex4}

\usepackage{amsmath,bm,amsfonts,amssymb}
\usepackage{graphics,graphicx}

\begin{document}

\title{Relaxation of Hot Quasiparticles in a ${\bm{d}}$-Wave Superconductor}

\author{P. C. Howell}
\author{A. Rosch}
%\email[]{Your e-mail address}
%\homepage[]{Your web page}
%\thanks{}
%\altaffiliation{}
\affiliation{\mbox{Institut f\"ur Theorie der Kondensierten Materie,
    Universit\"at  Karlsruhe, 76128 Karlsruhe, Germany}}
\author{P. J. Hirschfeld}

\affiliation{Department of Physics, University of Florida,
Gainesville, FL  32611-8440, USA }
\date{\today}

\begin{abstract}
Motivated by recent pump--probe experiments we consider the processes
by which ``hot'' quasiparticles produced near the antinodes of a
$d$-wave superconductor can relax. We show that in a large region of
momentum space processes which break Cooper pairs are forbidden by
energy and momentum conservation.  Equilibration
then occurs by scattering with thermal quasiparticles: 
Um\-klapp scattering is exponentially suppressed at low temperatures, but
small-angle scattering leads to power-law behavior.  By solving the
Boltzmann equation analytically we make detailed predictions for the
temperature and intensity dependence of these processes, which we
compare with experiment.
\end{abstract}
% insert suggested PACS numbers in braces on next line
% 74.25.Gz Optical properties 
% 74.72.-h Cuprate superconductors (high-Tc and insulating parent compounds) 
% 78.47.+p Time-resolved optical spectroscopies and other ultrafast optical
%  measurements in condensed matter 


\maketitle

{\it Introduction.}---Despite intensive effort for more
than a decade, no consensus exists on the form and origin of the
effective electron--electron interaction leading to pairing in the
cuprate superconductors.  Furthermore, although it is generally
agreed that the superconducting state has $d$-wave symmetry, many
properties of the Bogoliubov quasiparticle (QP) excitations 
are not well understood \cite{hussey02}. In conventional
superconductors, the study of non-equilibrium QP relaxation was used
successfully to extract information on residual particle--particle
interactions, as well as to pin down QP and phonon lifetimes (for a
review, see Ref.~\cite{LangenbergLarkin}).  The typical
time-resolved experiment is a measurement of the change in the
system's dielectric constant as a function of time following a pump
pulse which creates a non-equilibrium QP distribution.  The excited
QPs decay to equilibrium over a series of timescales involving several
steps, including at least: (i) a cascade of pair production until a
quasi-equilibrium is reached between ``hot" QPs of roughly the gap
energy $\Delta_0$ and phonons of energy $2\Delta_0$, and (ii) slow
recombination of QPs into Cooper pairs. The timescales involved in
step (i) are ${\cal O}$(ps), but can be much longer in step (ii)
[${\cal O} ($ns--$\mu$s)] since energy is continually exchanged
between the electron and phonon systems until heat is removed at the
sample surfaces; this long decay is sometimes referred to as the
``phonon bottleneck".  These two rates were clearly identified in, for
example, recent pulsed synchrotron measurements on Pb \cite{carr00}.

{\it A priori}, several differences are to be expected in the cuprate
superconductors. The very strong interactions in the normal state and
the larger gap scale $\Delta_0$ suggest that electron--electron rather
than electron--phonon scattering is the dominant relaxation mechanism.
Furthermore, the gap anisotropy and the equilibration with thermal QPs
at the nodes leads to a new type of slow decay, which we call the
``antinodal bottleneck". This is the inability of antinodal QPs to
decay rapidly by pair-breaking while simultaneously satisfying energy
and momentum conservation, due to the large difference in velocities
at the nodal and antinodal points of the Fermi surface. (This idea has
also been proposed by Orenstein~\cite{Orensteinprivate}.)

Although several recent experiments on the cuprates have probed the
decay of non-equilibrium QP populations in time
\cite{feenstra97,kabanov99,kaindl00,averitt01,schneider02,segre02,gedik03b,gedik03},
their methods and results differ considerably and it is not clear yet
whether a consistent interpretation can be extracted.  A short
relaxation process ${\cal O}(1$--10 ps) has indeed been
identified~\cite{kabanov99,kaindl00,averitt01,schneider02} in
experiments with pump pulse energy much greater than the gap scale.
In addition, a much longer component ${\cal O}(10\,\mu$s) component
has been observed \cite{feenstra97} and identified with a
recombination bottleneck.  Recently, however, Segre {\it et al.\ 
}\cite{segre02} pointed out that the early relaxation component in
their measurements on ortho-II YBCO$_{6.52}$ single crystals was
strongly $T$ and intensity dependent, and in the limit of weak
intensity they fitted it to a $T^3$ law.  At low $T$, this ``slow"
component was of order hundreds of ps, but still many orders of
magnitude shorter than that seen in Ref.~\cite{feenstra97}.

Here we present a theory of the single aspect of the complicated
non-equilibrium physics of pump--probe experiments most peculiar to the
$d$-wave superconductor, namely the mechanism intermediate between
steps (i) and (ii) whereby hot QPs scatter through the antinodal
bottleneck before recombination (processes on longer timescales have
been discussed in Ref.~\cite{nicol03}). Within a model where QPs
are scattered by a simple local interaction, we show that at high $T$
there is a fast relaxation due to Umklapp scattering, but below some
crossover temperature relaxation is dominated by diffusion in momentum
space along the Fermi surface from the antinode to the nodes. This
idea is a momentum-space analog of real-space ``QP traps" in
conventional superconductors~\cite{booth87}, in the sense that there
is an intermediate stage of relaxation in which QPs diffuse to a
region of lower gap.

\begin{figure}
\includegraphics[width=78mm]{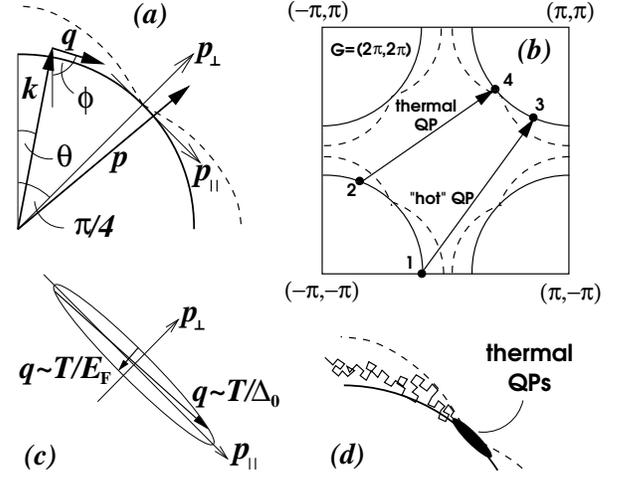}
\caption{\label{fig1}
  (a)~Coordinate systems for the QP momenta (heavy line: Fermi
  surface, dashed line: superconducting gap). Momenta $\bm{k}$ near
  the antinodes are given in polar coordinates $(k_{\text F}
  +k,\theta)$ and momenta $\bm{p}$ near the nodes in
  Cartesian coordinates $(p_\perp, p_\parallel)$ relative to the node.
  The antinodal QP scatters from $\bm{k}$ to $\bm{k+q}$, where
  $\bm{q}=(q,\phi)$ is in polar coordinates.  (b)~A typical Umklapp
  scattering process.  (c)~Typical magnitudes of the momentum transfer
  for scattering perpendicular and parallel to the nodal Fermi
  surface.  (d)~Schematic picture of diffusion of a hot antinodal QP
  towards the node.  }
\end{figure}

{\it Antinodal bottleneck.}---The (extrapolated) relaxation rate of
QPs of energy $\sim$$\Delta_0$ vanishes as $T \rightarrow 0$ in the
experiment of Segre {\it et al.\ }\cite{segre02}, implying both that
the Ortho II sample is extremely pure, so that impurities may be
neglected, and that pair-breaking processes are forbidden. We now show
that the latter effect is a simple consequence of the requirements of
energy and momentum conservation. We assume a $d$-wave order parameter
$\Delta_{\bm k}=\Delta_0\cos 2\theta$ over a Fermi surface
parameterized by $\theta$. In the coordinate systems of
Fig.~\ref{fig1}(a) the BCS dispersion relation can then be expanded
for momenta $\bm k$ near the antinode as $ E_{\bm{k}} = \Delta_{\bm k}
+ 2k^2E_\text{F}^2/\Delta_{\bm k}$ and for momenta $\bm p$ near the
node as $E_{\bm{p}} = 2 \sqrt{E_\text{F}^2p_\perp^2 + \Delta_0^2
p_\parallel^2 }$ (with $k_\text{F}$=1).

For simplicity, consider an antinodal QP exactly on the Fermi surface,
which for small transfered momentum ${\bm q}$ must scatter parallel to
the Fermi surface, $\phi \approx \frac{\pi}{2} - \theta$ [see
Fig.~\ref{fig1}(a)]; the energy lost is then $E_\text{lost} \approx
2q\Delta_0 \sin 2\theta$. The minimum energy required to break a
Cooper pair, {\it i.e.,~}to create two nodal QPs with momenta $\bm{p}$
and $-\bm{p} - \bm{q}$, is $2qE_\text{F}\sin|\frac{\pi}{4}-\theta|$,
which for $|\frac{\pi}{4}-\theta|>\Delta_0/E_\text{F}$ is larger than
$E_\text{lost}$.  More generally, pair-breaking cannot occur for $k
\lesssim 1/\sqrt{2} (\Delta_0/E_\text{F})^2 \cos^2 2\theta$, where $k$
is the distance from the Fermi surface. The origin of this bottleneck
is that the velocity of the antinodal QP is much less than nodal
velocities. While the above analysis neglects Umklapp scattering,
including it only modifies the picture quantitatively: the bottleneck
for pair-breaking remains but its size now depends on details of the
Fermi surface.

How, then, can a hot QP relax, trapped as it is in the bottleneck? It
could be scattered by phonons, impurities, various electronic
collective excitations, thermal (nodal) QPs or other hot (antinodal)
QPs.  For the creation of phonons there is an analogous bottleneck,
although scattering from thermally excited phonons is possible.  We
now investigate in detail the consequences of scattering from thermal
QPs; contributions from other hot QPs can be separated by their
dependence on the laser intensity (see below).  While our calculation
includes only QP--QP scattering, we expect qualitatively similar
results for scattering from (non-critical) spin fluctuations, as has
been shown to be the case for the optical conductivity up to a
phenomenological prefactor~\cite{walker00,duffy01}.

In the spirit of a Fermi liquid approach we assume the existence of
well-defined QPs, $d^\dag_{\bm{k}\uparrow} = u_{\bm{k}}
c^\dag_{\bm{k}\uparrow} - 
v_{\bm{k}} c_{-\bm{k}\downarrow}$, where $u_{\bm{k}}$ and $v_{\bm{k}}$
are the usual Bogoliubov coefficients. At low $T$ two-particle
processes dominate, parameterized by the following Hamiltonian:
%\begin{widetext}
%\begin{equation}
%  {\cal H} = \sum \Big[ 
%  \begin{aligned}[t]
%  &V_{2-3}^\text{pp} r_{12} r_{34} d^\dag_{4\uparrow} 
%    d^\dag_{3\downarrow} d^\dag_{2\downarrow} d^\dag_{1\uparrow} 
%    + V_{2+1}^\text{pb} \tilde{r}_{12} r_{34} d^\dag_{4\uparrow}
%    d^\dag_{3\downarrow} d^\dag_{2\sigma} d_{1\sigma} 
%    + V_{2+1}^\text{qq} r_{12} r_{34} d^\dag_{4\sigma}
%    d^\dag_{3\bar{\sigma}} d_{2\bar{\sigma}}d_{1\sigma} \\
%  &+ V_{2-3}^\text{qq} \tilde{r}_{14} \tilde{r}_{23} d^\dag_{4\sigma}
%  d^\dag_{3\sigma'} d_{2\sigma'} d_{1\sigma} 
%  + \text{H.c.} \Big] 
%\;,  \label{hamiltonian}
%\end{aligned}
%\end{equation}
%\end{widetext}
\begin{equation}
  {\cal H} = \sum \Big[ 
  \begin{aligned}[t]
  &V_{2-3}^\text{pp} r_{12} r_{34} d^\dag_{4\uparrow} 
    d^\dag_{3\downarrow} d^\dag_{2\downarrow} d^\dag_{1\uparrow} \\
  &+ V_{2+1}^\text{pb} \tilde{r}_{12} r_{34} d^\dag_{4\uparrow}
    d^\dag_{3\downarrow} d^\dag_{2\sigma} d_{1\sigma} \\
  &+ V_{2+1}^\text{qq} r_{12} r_{34} d^\dag_{4\sigma}
    d^\dag_{3\bar{\sigma}} d_{2\bar{\sigma}}d_{1\sigma} \\
  &+ V_{2-3}^\text{qq} \tilde{r}_{14} \tilde{r}_{23} d^\dag_{4\sigma}
  d^\dag_{3\sigma'} d_{2\sigma'} d_{1\sigma} 
  + \text{H.c.} \Big] 
\;,  \label{hamiltonian}
\end{aligned}
\end{equation}
where $i \equiv \bm{k}_i, r_{ij}=u_iv_j+v_iu_j,
\tilde{r}_{ij}=u_iu_j-v_iv_j$ and the sum conserves crystal momentum.
We assume that the effective interactions $V$ are only weakly
dependent on momentum (this is not justified if long-ranged
spin-fluctuations are important).  Hot QPs predominantly relax by
scattering from thermal QPs, as expressed by the last two terms;
energy conservation forbids leading contributions from the first term,
while the bottleneck discussed above blocks pair-breaking processes
from the second term.  The BCS coherence factors $r_{ij},
\tilde{r}_{ij}$ are of order unity for the relevant processes and so
will be neglected.

There are two types of scattering of hot QPs from thermal QPs,
distinguished by their typical momentum transfer.  We first
discuss Umklapp scattering, which involves large transfered
momenta, $q \sim k_\text{F}$, as illustrated in Fig.~\ref{fig1}(b).
A hot QP undergoing Umklapp and subsequent pair-breaking scattering
can reach the node in a small number of scattering events.  However,
due to kinematic constraints such scatterings are only possible if
the thermal QP is some minimum angle away from the node and therefore
has a minimum energy $\Delta_\text{U}$. Accordingly, the corresponding
scattering rate is exponentially suppressed. Using Fermi's golden rule
we obtain
\begin{equation}
\frac{1}{\tau_\text{U}} \approx M_\text{qq}
\frac{(2\pi\Delta_\text{U} T)^{3/2}}{\Delta_0/E_\text{F}}
  \,\text{e}^{-\Delta_\text{U}/T}
  \,,\; M_\text{qq}=
  \frac{2\pi V_\text{qq}^2 a^4}{\hbar(2\pi \hbar v_\text{F})^4}\,,
\label{Umklapprate}
\end{equation}
where $a$ is the lattice constant, $v_\text{F}$ is the Fermi velocity
and the size of $\Delta_\text{U} < \Delta_0$ depends on the details of
the Fermi surface shape. An analogous Umklapp gap determines the
momentum relaxation of nodal QPs~\cite{walker00} and hence the optical
conductivity at low frequency.

At low $T$ the second type of process becomes important: normal
scattering, which we now show involves small momenta, $q \ll k_F$. For
a given $\bm q$ the slow anti\-nodal QP loses energy much less than
$q\Delta_0$, so that the energy of the fast nodal QP changes little,
$E_{\bm p} \approx E_{\bm{p-q}} $.  Therefore the typical $q$ depends
on direction [see Fig.~\ref{fig1}(c)], and is $q \approx 2p_\| \sim
T/\Delta_0$ for scattering parallel to the Fermi surface at the nodes
($\phi \approx \frac{\pi}{4}$ and symmetry-related angles), and $q
\approx 2p_\perp \sim T/E_\text{F}$ in perpendicular directions.

Since $q\ll k_\text{F}$, the hot QP experiences many collisions with
the nodal QPs before it reaches the nodes, as sketched in
Fig.~\ref{fig1}(d). This has two important consequences: (i)~the
trajectory of the hot QP is diffusive in momentum space, and (ii)~the
hot QP thermalizes in the radial direction with the nodal QPs (see
below). The typical momentum of a hot QP is thus determined by
$E_{\bm{k}} -\Delta_0\cos2\theta \sim T$ and so $k \sim k_\text{th} =
\sqrt{T\Delta_0}/E_\text{F}$.

{\it Boltzmann equation.}---To calculate the time $\tau_\theta$ for a
dilute distribution $g_{\bm{k}}(t)$ of antinodal QPs to reach the
nodes we first consider the situation where the radial distribution of
hot QPs lies entirely within the bottleneck, $k_\text{th} \ll
(\Delta_0/E_\text{F})^2$, which requires
$T\ll\Delta_0^3/E_\text{F}^2$. We assume the nodal QPs are always in
local thermal equilibrium, as $\tau_\theta^{-1}$ will turn out to be
much slower than the nodal relaxation rate $\sim$$M_\text{qq}
T^3E_\text{F}^2/\Delta_0^2$. Linearizing in $g_{\bm k}$ the Boltzmann
equation reads
\begin{gather}
\!\!  \frac{\partial g_{\bm k}}{\partial t} = \int \text{d}\bm{q} \Big[
 g_{\bm{k}-\bm{q}} G_\text{qq}(\bm{q}, \varepsilon_{\bm{k}-\bm{q},\bm{q}})
 - g_{\bm k} G_\text{qq}(\bm{q}, \varepsilon_{\bm{k},\bm{q}})
  \Big] \,, \\
  G_\text{qq}(\bm{q},\varepsilon)
  = M_\text{qq} \int \text{d}\bm{p} \,f_{\bm p} (1-f_{\bm{p}-\bm{q}})
  \,\delta(\varepsilon_{\bm{p},-\bm{q}} +\varepsilon) \;.
\end{gather}
Here $f_{\bm p}$ is a Fermi function, $\varepsilon_{\bm{k},\bm{q}} =
E_{\bm k} - E_{\bm{k}+\bm{q}}$ is the energy transfer, and
$G_\text{qq}$ is the integral over momenta near all four nodes of
QP--QP scattering processes.

Since we expect the hot QPs to thermalize in the radial direction we
make the transformation
\begin{equation}
g_{\bm k}(t) = \frac{C_{\bm k}(t)}{h_{\theta_{\bm k}}} 
  \,\text{e}^{-\beta E_{\bm k}}
\;,\; 
h_{\theta_{\bm k}} = \int (1+k)\text{d}k \, \text{e}^{-\beta E_{\bm k}} \;,
\label{gAnsatz}
\end{equation}
where $C_{\bm k}(t)$ is a smooth function of $\bm k$ to be determined.
[The factor of $(1+k)$ comes from the use of polar coordinates.]
Using detailed balance and \eqref{gAnsatz}, the Boltzmann equation now
takes the form
\begin{eqnarray}
\frac{1}{h_{\theta_{\bm k}}} \,\frac{\partial C_{\bm k}}{\partial t}
  = \int \text{d}\bm{q}
  \,G({\bm q},\varepsilon_{\bm{k},\bm{q}})
  \left[ \frac{C_{\bm{k}+\bm{q}}}{h_{\theta_{\bm{k}+\bm{q}}} } -
    \frac{C_{\bm k}}{h_{\theta_{\bm k}}} \right] .
\label{boltz2}
\end{eqnarray}

If the hot QPs are close to {\it radial} equilibrium we expect $C_{\bm
  k}$ to depend only weakly on $k$ and therefore expand it as $C_{\bm
  k}(t) = C^0_{\theta_{\bm k}}(t) + (k-\langle k\rangle)\,
C^1_{\theta_{\bm k}}(t) + \cdots$, where $\langle k \rangle =
h_{\theta_{\bm k}}^{-1} \int (1+k) {\rm d}k \,k \,{\rm e}^{-\beta
  E_{\bm k}}$ such that $\int \text{d}\bm{k} \,g_{\bm k}(t) = \int
\text{d}\theta \,C^0_\theta(t)$.  The equation for $C^0_\theta$ can
then be projected out from \eqref{boltz2} by integrating over radial
directions. Since $q \,\partial_\theta(C^0_\theta/h_\theta) \ll
C^0_\theta/h_\theta$ we can perform a gradient expansion, leading to
(for $\theta \ll \smash{ \frac{\pi}{4} }$) 
\begin{equation}
\frac{\partial}{\partial t} \left[\frac{C^0_\theta(t)}{h_\theta}\right]
  = \left[ F\theta \,\frac{\partial}{\partial\theta}
  +D \,\frac{\partial^2}{\partial\theta^2} \right]
  \frac{C^0_\theta(t)}{h_\theta}
  \;, \label{diffuseeq}
\end{equation}
\begin{eqnarray}
F = \frac{1}{\theta} \int \text{d}^2\bm{q} \, \text{d}k
  \,\frac{\text{e}^{-\beta E_{\bm k}}}{h_\theta} \,q \sin(\theta+\phi)
  \,G_\text{qq}(\bm{q}, \varepsilon_{\bm{k},\bm{q}})
  \;, \label{Fdef} \\
D = \int \text{d}^2\bm{q} \, \text{d}k
  \, \frac{\text{e}^{-\beta E_{\bm k}}}{h_\theta}
  \frac{q^2\sin^2(\theta+\phi)}{2(1+k)}
  \, G_\text{qq}(\bm{q}, \varepsilon_{\bm{k},\bm{q}})
  \;. \label{Ddef}
\end{eqnarray}
The main contribution is from $\phi \approx \frac{\pi}{4}$ leading to
``zig-zag'' diffusion in momentum space [see Fig.~\ref{fig1}(d)], and
hence
\begin{equation}
F \approx \frac{7\pi^5}{80} \,\frac{M_{\text{qq}}T^4E_\text{F}^2}{\Delta_0^3}
\;,\quad D= F\frac{T}{4\Delta_0} \;.
\end{equation}
We have dropped terms in $C^i_\theta \,(i\geq 1)$ on the right-hand
side of \eqref{diffuseeq}
since they decay faster than $C^0_\theta$ (see below).

As $F/D = 4\Delta_0/T$  we can rewrite
\eqref{diffuseeq} for $\theta \ll \frac{\pi}{4}$ as
\begin{equation}
\partial_t \,C^0_\theta = -F \,\partial_\theta (\theta C^0_\theta)
+ D \,\partial^2_\theta C^0_\theta \;.
\label{diffuseeq2}
\end{equation}
Now $D$ can be identified as the diffusion constant in momentum space,
while $F$ parameterizes the force driving hot QPs towards the nodes.
We emphasize that the forms of \eqref{diffuseeq} and
\eqref{diffuseeq2} are completely determined by detailed balance,
local conservation of the number of hot QPs and symmetry under $\theta
\to -\theta$.  The Boltzmann distribution $C^0_\theta \propto
h_\theta$ obviously solves \eqref{diffuseeq}, while the conservation
of the number of hot QPs, $N_\text{hot} = \int \text{d}\theta
\,C^0_\theta(t)$, is manifest in \eqref{diffuseeq2}; the equivalence
of \eqref{diffuseeq} and \eqref{diffuseeq2} fixes $F/D$.

For an initial Gaussian distribution of width $\alpha$ the solution to
\eqref{diffuseeq2} is a Gaussian of width $\sigma(t)$, where
$\sigma^2(t)= (2D/F+\alpha^2) \text{e}^{2Ft}- 2D/F $.  For short
times, $t\ll 1/F$, the motion is diffusive, $\sigma = 2\sqrt{D t}$,
while for $t\gg 1/F$ we obtain $\sigma \propto {\rm e}^{F t}$.  The
antinodal QPs reach the node after a time $\tau_\theta$ such that
$\sigma(\tau_\theta)\sim \pi/4$, and hence
\begin{eqnarray}
  1/\tau_\theta \sim F/\ln(\min [ \sqrt{F/D},\alpha^{-1} ])
  \sim M_\text{qq} T^4 E_\text{F}^2/\Delta_0^3
\;. \label{lowTrate}
\end{eqnarray}
Details of the initial distribution
only give log corrections.

To check the assumption of radial equilibrium we calculate
perturbatively the evolution of the corrections $C_\theta^1(t),
C_\theta^2(t) \ldots$ using \eqref{boltz2}. Deviations from radial
equilibrium decay rapidly, $\partial C^i_\theta/\partial t \approx -
C^i_\theta/\tau_i$ where $\tau_i \lesssim
\tau_\theta \Delta_0^2/E_\text{F}^2\ll \tau_\theta $ for $i\ge 1$, and
can therefore be neglected.

{\it Higher temperatures.}---We now consider higher temperatures,
$\Delta_0^3/E_\text{F}^2\ll T \ll\Delta_0^2/E_\text{F}$, where the
upper limit comes from the neglect of curvature in the nodal
dispersion $E_{\bm p}$. In this regime the typical antinodal velocity
$4k_\text{th}E_\text{F}^2/\Delta_0 = 4E_\text{F}\sqrt{T/\Delta_0}$ is
now large compared to the nodal velocity parallel to the Fermi
surface, but still small compared to the perpendicular nodal velocity.
This has two consequences. (i)~Pair-breaking (and recombination) is
now allowed, as $k_\text{th} \gg \Delta_0^2/E_\text{F}^2$,
{\em i.e.~}the QP is outside the bottleneck;
in such a process one QP turns into three, of which two must
reside at the nodes (due to energy and momentum conservation).
(ii)~Typical transfered momenta still
turn to be small, for both pair-breaking and QP--QP processes,
although now
$q \sim \sqrt{T\Delta_0}/E_\text{F}$ for scattering parallel to the nodal
Fermi surface (instead of $q \sim T/\Delta_0$ when
$T\ll\Delta_0^3/E_\text{F}^2$).

All physical requirements which led to the form of \eqref{diffuseeq},
\eqref{diffuseeq2} and fixed the ratio $F/D$ are therefore fulfilled:
conservation of {\em antinodal} QPs, small momentum transfer and
detailed balance (since both pair-breaking and recombination are
possible).  Generalizing \eqref{Fdef}, we find
\begin{equation}
  1/\tau_\theta  \approx \big(0.92 M_\text{qq} + 0.67 M_\text{pb}\big)
  \sqrt{T^5\Delta_0^3}/E_\text{F} \;, \label{highTrate}
\end{equation}
with $M_\text{pb}$ defined analogously to $M_\text{qq}$,
Eq.~(\ref{Umklapprate}).

{\it Discussion.}---We have shown that for vanishing laser intensity
there are two relaxation mechanisms for hot QPs: Umklapp scattering
from thermal QPs with an exponential relaxation rate
\eqref{Umklapprate}, and normal scattering with a power-law rate
(\ref{lowTrate},\,\ref{highTrate}).  While early experiments were
fitted to a $T^3$ law~\cite{segre02}, later
results~\cite{Orensteinprivate,gedik03b} suggest activ\-ated behavior
for the lowest measured temperature (down to 10\,K), $\tau^{-1}
\approx \gamma\,\text{e}^{-E_\text{act}/T}$, with
$\gamma\approx8\times 10^{12}$\,s$^{-1}, E_\text{act} \approx 100$\,K.
This is consistent with the picture given above, where from
\eqref{Umklapprate} and \eqref{highTrate} the crossover to $T^{5/2}$
behavior is expected to occur when $\text{e}^{-E_\text{act}/T} \sim
T\Delta_0/E_\text{F}^2$, or $T\approx 8$\,K, assuming $\Delta_0
\approx 200$\,K~\cite{dagan00} and $E_\text{F} \approx
8\Delta_0$~\cite{sutherland03}. The surprisingly low crossover scale
$\sim$$0.1E_\text{act}$ arises since many small-momentum scattering
events are required for the hot QP to reach the nodes and therefore
\eqref{highTrate} is small. Note that the crossover from
$T^{5/2}$~\eqref{highTrate} to $T^4$~\eqref{lowTrate} is expected
around $T \sim \Delta_0^3/E_\text{F}^2 \approx 1$\,K.

As a consistency check of our approach we now consider the
absolute  magnitude of the relaxation rate,  for which we require
$M_\text{qq}$ in \eqref{Umklapprate}.  Assuming that a similar
matrix element appears in the optical conductivity, we can use the
fit of Walker and Smith~\cite{walker00} to YBCO$_{6.99}$ to find
$M_\text{qq} \approx 2\times 10^{17}$\,eV$^{-3}$\,s$^{-1}$. This
gives $\tau_\text{U} \approx 1$\,ns at $T=20$\,K, to be compared
to the experimental value of 0.03\,ns. Given the possible
momentum-dependence of $M_\text{qq}$, Fermi surface shape effects,
the difference in doping in the two experiments and especially the
uncertainty in $E_\text{act}$, this is not unreasonable agreement.

Within our model we can also estimate the intensity dependence of the
scattering rate, which experimentally~\cite{segre02,gedik03b} is
given by $\tau_n^{-1} \approx c\rho$ almost independent of $T$, where
$\rho$ is the energy density deposited by the laser and $c\approx 1.2
\times 10^6~{\rm s^{-1}J^{-1}m^3}$.  Assuming a uniform distribution
of hot QPs on the Fermi surface, we use Fermi's golden rule to find
$1/\tau_{n}\approx 8\pi M_\text{qq}n_\text{hot} (\hbar
v_\text{F})^4\ln(E_\text{F}/\Delta_0)\Delta_0/E_\text{F}^2$, where
$n_\text{hot}$ is the (planar) density of hot QPs.  With the above
numbers, a comparison to the experiment would suggest that only
$\sim\!0.1\%$ of the deposited energy is used to create $n_\text{hot}$
hot QPs each of energy $\sim$$\Delta_0$; equivalently, assuming
$100\%$ energy conversion, our predicted rates are a factor $10^3$ too
large.  While a sizable fraction of the laser energy probably goes
into the phonon system, we think this is unlikely to explain the
discrepancy completely.

{\it Conclusions.}---Using BCS theory for $d$-wave superconductors we
have calculated the electronic contribution to the relaxation of hot
QPs.  An analytical solution to the QP Boltzmann equation shows that
small momentum scattering leads to a diffusive dynamics of antinodal
QPs and associated power-law relaxation rates, unfortunately for $T<
10\,$K which has not yet been investigated experimentally. At higher
$T$ the relaxation is dominated by Um\-klapp scattering with large
momentum transfer leading to exponential $T$ dependence, consistent
with experiments on ultra-clean samples.  However, it is currently
unclear whether our estimates for magnitudes of relaxation times are
correct, and improvement will require a more complete understanding of
the relation between the incident photon intensity and the number of
hot QPs.

There are still many details and apparent discrepancies between the
various non-equilibrium measurements on the cuprates which we cannot
understand at present, and it is clear that more realistic physics
must be incorporated to describe the evolution of the QP distribution
over all timescales of interest, at different dopings, and in dirtier
samples, as well as the dynamics of the order parameter. We believe
nevertheless that the theoretical analysis of these experiments is
worth pursuing since they potentially provide a window on the
frequency and momentum dependence of the interactions responsible for
superconductivity in these systems.

{\it Acknowledgements.} This work was supported by the Emmy Noether
programme of the DFG and NSF-DMR 99-74396. The authors thank
M.~Eschrig, N.~Gedik, R. Kaindl, P.~Lemmens, J.~Orenstein, D.~J.
Scalapino, M.~Vojta and P.~W\"olfle for stimulating conversations.


\begin{thebibliography}{99}

\bibitem{hussey02} N. E. Hussey, Adv. Phys. {\bf 51}, 1685 (2002).

%\bibitem{Gray}K. E. Gray,  {\it Nonequilibrium Superconductivity, Phonons and
%Kapitza Boundaries}, ed. K. E. Gray (New York: Plenum Press), 1981.

\bibitem{LangenbergLarkin}{\it Nonequilibrium Superconductivity},
edited by D. N. Langenberg and A. I. Larkin (North-Holland, New York, 1986).

%\bibitem{ChangScalapinoReview} J. J. Chang and D. J. Scalapino, in
%{\it Superconductor Applications: Squids and Machines (1977)},
%eds. B. B. Schwartz and S. Foner, (New York: Plenum Press), 1977.

%\bibitem{carr00}G. L. Carr, R. P. S. M. Lobo, J. LaVeigne,
%  D. H. Reitze, and D. B. Tanner, Phys. Rev. Lett. {\bf 85}, 3001 (2000).

%\bibitem{feenstra97}B. J. Feenstra, J. Schutzmann, D. van der
%Marel, R. Perez Pinaya, and M. Decroux, Phys. Rev. Lett. {\bf 79}, 4890
%(1997).

%\bibitem{kabanov99} V. V. Kabanov, J. Demsar, B. Bodobnik, and D.
%Mihailovic, Phys. Rev. B {\bf 59}, 1497 (1999).

%\bibitem{kaindl00}R. A. Kaindl, M. Woerner, T. Elsaesser, D. C. Smith, J. F.
%Ryan, G. A. Farnan, M. P. McCurry, and D. G. Walmsley, Science {\bf 287},
%470 (2000).

%\bibitem{averitt01} R. D. Averitt, G. Rodriguez, A. I. Lobad, J. L. W.
%Siders, S. A. Trugman, and A. J. Taylor, Phys. Rev. B {\bf 63}, 140592
%(2001).

%\bibitem{segre02} G. P. Segre, N. Gedik, J. Orenstein, D. A. Bonn,
%Ruixing Liang, and W. N. Hardy, Phys. Rev. Lett. {\bf 88}, 137001
%(2002).

%\bibitem{gedik03} N.~Gedik, J.~Orenstein, Ruixing Liang, D.~A. Bonn
%and W.~N. Hardy, Science {\bf 300} 1410 (2003).

\bibitem{carr00}G. L. Carr {\em et al.}, Phys. Rev. Lett. {\bf 85}, 3001 (2000).

\bibitem{Orensteinprivate} J.~Orenstein (private communication).


\bibitem{feenstra97}B. J. Feenstra {\em et al.}, Phys. Rev. Lett. {\bf 79}, 4890
(1997).

\bibitem{kabanov99} V. V. Kabanov {\em et al.}, Phys. Rev. B {\bf 59}, 1497 (1999).

\bibitem{kaindl00}R. A. Kaindl {\em et al.}, Science {\bf 287},
470 (2000).

\bibitem{averitt01} R. D. Averitt {\em et al.}, Phys. Rev. B {\bf 63}, 140502
(2001).

\bibitem{schneider02} M. L. Schneider {\em et al.},
  Europhys. Lett. {\bf 60}, 460 (2002).

\bibitem{segre02} G. P. Segre {\em et al.}, Phys. Rev. Lett. {\bf 88},
  137001 (2002); J.~Demsar {\em et al.}, Phys. Rev. Lett. {\bf 91}, 169701
  (2003); N.~Gedik {\em et al.}, Phys. Rev. Lett. {\bf 91}, 169702
  (2003).

\bibitem{gedik03b} N. Gedik {\it et al.}, preprint cond-mat/0309121.

\bibitem{gedik03} N.~Gedik {\em et al.}, Science {\bf 300} 1410 (2003).

\bibitem{nicol03} E.~J. Nicol and J.~P. Carbotte, Phys. Rev. B, 
  {\bf 67} 214506 (2003).

\bibitem{booth87} N. E. Booth, Appl. Phys. Lett. {\bf 50}, 293 (1987).

\bibitem{duffy01} D. Duffy, P. J. Hirschfeld and D. J. Scalapino,
  Phys. Rev. B {\bf 64}, 224522 (2001).

\bibitem{walker00} M. B. Walker and M. F. Smith,  Phys. Rev. B {\bf 61},
11285 (2000).

%\bibitem{QSB94} S. M. Quinlan, D. J. Scalapino, and N. Bulut,  Phys. Rev. B {\bf 49}, 1470 (1994).

%\bibitem{Yas95} M. L. Titov, A. G. Yashenkin, and D. N. Aristov,
%Phys. Rev. B {\bf 52}, 10626 (1995).


%\bibitem{Zhao99} Y. G. Zhao, W. L. Cao, J. J. Li, H. D. Drew,
%  R. Shreekala, C. H. Lee, S. P. Pai, M. Rajeswari, S. B. Ogale,
%  R. P. Sharma, G. Baskaran and T. Venkatesan,
%J. Supercon. {\bf 12}, 675 (1999).

\bibitem{dagan00} Y.~Dagan, R.~Krupke and G.~Deutscher, Phys. Rev. B
  {\bf 62}, 146 (2000).

%\bibitem{sutherland03} M.~Sutherland, D.~G. Hawthorn, R.~W. Hill,
%  F.~Ronning, S.~Wakimoto, H.~Zhang, C.~Proust, E.~Boaknin, C.~Lupien,
%  L.~Taillefer, R.~Liang, D.~A. Bonn, W.~N. Hardy, R.~Gagnon,
%  N.~E. Hussey, T.~Kimura, M.~Nohara and H.~Takagi, Phys. Rev. B {\bf
%  67}, 174520 (2003).
\bibitem{sutherland03} M.~Sutherland {\em et al.}, Phys. Rev. B 
  {\bf  67}, 174520 (2003).
\end{thebibliography}
\end{document}